\documentclass[aps,twocolumn,showpacs]{revtex4}
\usepackage{graphicx}

\def\be7pg{$^7Be(p,\gamma)^8B$}
\def\xbe7{$^7Be$}
\def\b8{$^8B$}
\def\S17{$S_{17}(0)$}
\def\xs17{$S_{17}$}
\def\s34{$S_{34}(0)$}
\def\xpm{$\pm$}

\begin{document}

\title{How Accurately Do We Know the Formation of Solar $^8B$?}
\thanks{Work Supported by USDOE Grant No. DE-FG02-94ER40870.}

\author{Moshe Gai}
\altaffiliation{Permanent Address: Laboratory for Nuclear Science, Department of Physics, 
University of Connecticut, 2152 Hillside Rd., U3046, Storrs, CT 06269-3046.}
\affiliation{Department of Physics, Room WNSL102, Yale University, \\
PO Box 208124, 272 Whitney Avenue, New Haven, CT 06520-8124. \\
\    \\
e-mail: moshe.gai@yale.edu, URL: http://www.phys.uconn.edu}

\begin{abstract}

The large value of \S17 = 22.1 \xpm \ 0.6 eV-b, reported by the Seattle group, 
suggests a larger total \b8 solar neutrino flux. Together with the two high precision values quoted 
for \s34 it is either 20\% or 9\% larger than measured by SNO. While the accuracy of the Standard Solar 
Model has recently been revisited, precise nuclear inputs are still relevant, but a detailed examination 
of current data on \xs17 (as opposed to an examination of \S17 only) excludes quoting \S17 with sufficiently 
small uncertainty. In contrast to suggestions that \S17 is now known with the (impressive) accuracy of \xpm 3\%, 
the exact value of \S17 is dependent on the choice of the data and the choice of theory used for extrapolation. 
In addition recent high precision results (including the Seattle data) on \xs17 which are in good agreement, still 
differ on the measured slopes, as does the theory, precluding an accurate extrapolation to zero energy of the 
consistent data. Using a common extrapolation of only the consistent high precision data, suggests a value of 
\S17 = 21.2 \xpm \ 0.5 eV-b, but a value equal to or smaller than 19.0 eV-b can not be excluded due to 
the uncertainty in the extrapolation, leading to an additional error of $^{+0.0}_{-3.0}$ eV-b. A proposal 
to remedy this situation is discussed. 

\end{abstract}

\pacs{25.20.Dc, 25.70.De, 95.30-K, 26.30.+K, 26.65.+t}

\maketitle

The high precision measurement of neutral current interactions of "\b8 
solar neutrinos" in the SNO detector (with added salt) \cite{SNO}, yields the measured flux: 
$\phi_{NC}$ = 5.21 \xpm \ 0.27 (stat) \xpm \ 0.38 (syst) x $10^6$ cm$^{-2}$ sec$^{-1}$, 
with a precision of \xpm 5.2\% and accuracy of \xpm 7.3\%. A smaller total \b8 flux is deduced \cite{SNO} 
with the constraint of an undistorted \b8 energy spectrum. A global analysis of all solar experiments 
and KAMLAND yields the total \b8 solar neutrino flux with a precision of \xpm 4\% \cite{RoadMap}. 
It therefore mandates measurements of nuclear inputs to the Standard Solar Model (SSM) \cite{SSM} with similar 
uncertainty of 5\%. In this paper we address the accuracy of our knowledge of the cross section for 
forming \b8 in the sun via the proton capture, \be7pg reaction.

Several new measurements of the cross section of the \be7pg reaction 
\cite{Iw99,Ham01,Str01,Dav01,Seatt,Weiz,Sch03} were reported after the publication by 
Adelberger {\em et al.} \cite{Adel}. If one adopts the value of the astrophysical cross 
section factor (S-factor as defined in \cite{Adel}) measured by the Seattle group 
\cite{Seatt}, \S17 = 22.1 \xpm \ 0.6 eV-b, than it increases by 16\% the prediction of the 
SSM that assumes \S17 = 19.0 eV-b for a total \b8 neutrino flux of 
5.87 x $10^6$ cm$^{-2}$ sec$^{-1}$.
  
When adopting the high precision value of \S17 measured by the Seattle group, the uncertainty of 
the predicted total \b8 flux due to nuclear inputs is no longer dominated by \S17. 
Instead it is dominated by the two values found for \s34 \cite{Adel}. Note that each value of 
\s34 is however known with 3-4\% uncertainty \cite{Adel}, hence propagating to a small uncertainty 
of 2-3\% in the predicted total \b8 neutrino flux.

The Seattle large value of \S17 \cite{Seatt} together with the larger value of 
\s34 = 572 \xpm 26 eV-b deduced from \xbe7 activity measurements \cite{Adel,Ham} 
yield a predicted total \b8 neutrino flux that is 20\% larger than measured by SNO. The smaller 
value of \s34 = 507 \xpm 16 eV-b \cite{Adel} reduces the discrepancy to 9\%. 
The quoted average value \s34 = 530 \xpm 50 eV-b \cite{Adel} yield a discrepancy 
of 13\%.

The uncertainty of the predicted SSM flux arising from other than nuclear cross section factor considerations was 
quoted \cite{SSM} to be essentially \xpm 5\%, smaller than the uncertainty in the SNO measured flux 
of \xpm 7.3\%. Very recently the accuracy of the SSM was re-evaluated \cite{Bahcall} in view of a re-evaluation 
of the (surface) chemical composition of the sun (Z/X). It now appears that the  prediction of the \b8 flux has 
considerably larger uncertainty (\xpm 23\%), mostly due to the unknown chemical composition of the sun \cite{Bahcall}  
and an error bar of 6\% \cite{SSM} increased to 20\% \cite{Bahcall}. None-the-less the need still exists for 
high precision (\xpm 5\% or better) measurement of \S17, and most certainly of \s34, for example should one later 
improve the accuracy of our knowledge of the surface composition of the sun. And certainly the question whether this 
precision has been achieved is of interest in of itself. 

In contrast to several repeated statements that \S17 is now known with an (impressive) accuracy of 3\%, we demonstrate 
that the exact value of \S17 is dependent on the choice of data as well as the choice of the theory that one uses to 
extrapolate \S17. This situation can only be alleviated by new measurement(s) as we discuss below.

\begin{figure}
 \includegraphics[width=3in]{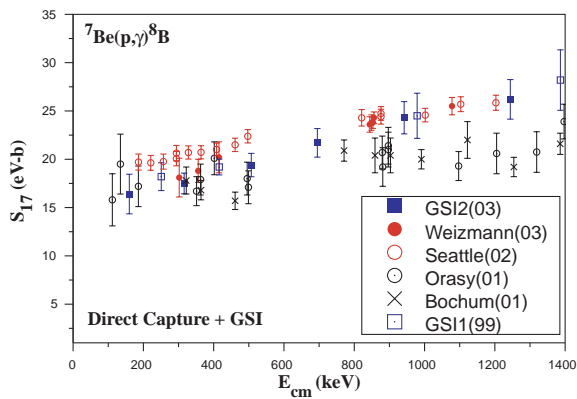}
 \caption{\label{World} Recent measured data on \xs17. The resonance data at 632 keV 
 measured in a few experiments are not shown.}
\end{figure}

The recent direct capture measurements of the $^7Be(p,\gamma)^8B$ reaction with 
$^7Be$ targets \cite{Ham01,Str01,Seatt,Weiz}, are shown in Fig. 1, together 
with the two most recent results of the Coulomb dissociation of \b8 measured at the 
GSI facility \cite{Iw99,Sch03}. We refer the reader to \cite{Adel} for a review 
of all previous measurements of the direct capture reaction and the Coulomb dissociation, 
as well as other indirect measurements of \S17 that are not relevant for the 
current discussion. The status of the world data on \xs17 shown in 
Fig. 1 is very unsatisfactory with specific measured data points differing by even 
more than 4 sigma. Hence inclusion of all data shown in Fig. 1 for extracting a value 
for \S17 necessarily leads to wrong conclusions.

In contrast the data of the Seattle \cite{Seatt}, 
Weizmann \cite{Weiz}, GSI1 \cite{Iw99} and GSI2 \cite{Sch03} 
groups, that were measured with high precision of the order of \xpm 3-5\%, 
exhibit a remarkably good agreement (with no data points more than 2 sigma away from the average). In sharp 
contrast the discrepant direct capture data \cite{Ham01,Str01} suggest large ill understood systematic 
differences. These direct capture data of the Orsay \cite{Ham01} and Bochum groups \cite{Str01} were measured 
with lower precision of the order of \xpm 10\%, and together with the Coulomb dissociation results of the MSU 
group \cite{Dav01} that will be discussed below, are not consistent with the above mentioned high 
precision measurements of Seattle, Weizmann, GSI1 and GSI2 
\cite{Seatt,Weiz,Iw99,Sch03}. These discrepant data can not be included in the same 
global analysis of \S17, unless the large systematic uncertainties are well understood 
and corrected.

\begin{figure}
 \includegraphics[width=3in]{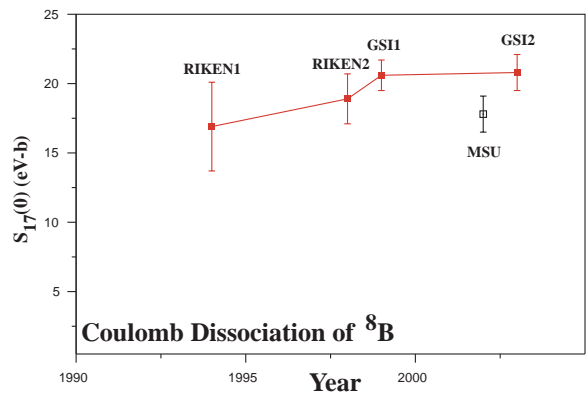}
 \caption{\label{CD} Extrapolated \S17 from data on the Coulomb dissociation of 
 \b8 using the theoretical calculation of Descouvemont and Baye \cite{DB}, as discussed 
 in the text}
\end{figure}

\begin{figure}
 \includegraphics[width=3in]{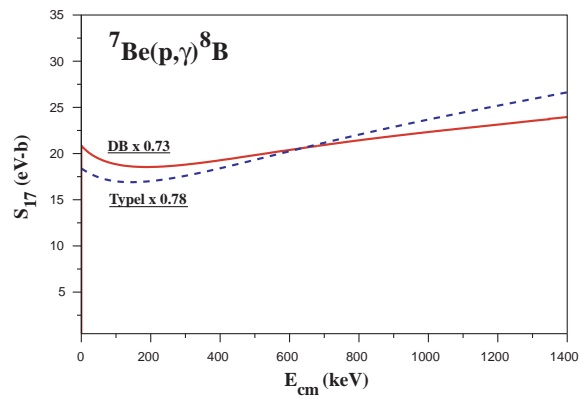}
 \caption{\label{Theory} The normalized theoretical calculations of 
\xs17 by Descouvemont and Baye \cite{DB}, and the potential model of 
Typel \cite{Sch03}. The normalization factors are shown.}
\end{figure}

\begin{figure}
 \includegraphics[width=3in]{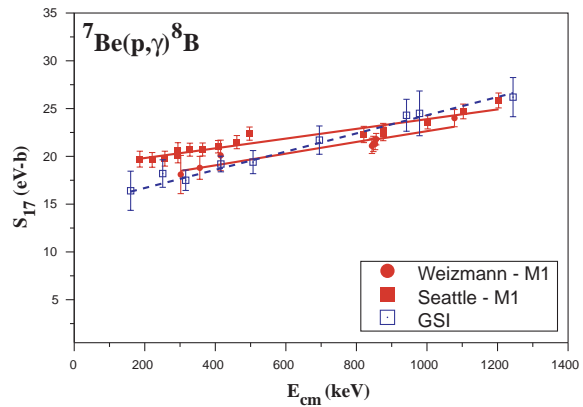}
 \caption{\label{Extrap} A Comparison of the recent Seattle(02) \cite{Seatt}, 
Weizmann(03) \cite{Weiz}, and GSI \cite{Iw99,Sch03} data. An M1 contribution due 
to the resonance at 632 keV is subtracted from the direct capture data.   
In spite of the good agreement, the measured slopes of the astrophysical 
cross section factor are sufficiently different, see Table II, precluding 
an accurate ($\pm$5\%) extrapolation to zero energy.}
\end{figure}

\begin{table}[hb] 

\caption[TableI]{Extrapolated cross section factors using the calculation of 
Descouvemont and Baye \cite{DB}. Only recent high precision results, \S17 
measured with an error of approximately \xpm 5\% or better, are shown, excluding 
the recent results of: RIKEN2(98) ($18.9 \pm 1.8$) \cite{Kik}, Orsay(01) 
($18.8 \pm 1.7$) \cite{Ham01} and Bochum(01) ($18.4 \pm 1.6$) 
\cite{Str01}, as discussed in the text.}\label{tab1}

\begin{tabular}{lllll}

\hline\\[-10pt]
Experiment & $S_17(0)$ (eV-b) \\
\hline\\[-10pt]
GSI1(99) \cite{Iw99} & $20.6 \ + 1.2 \ - 1.0$  \\
GSI2(03) \cite{Sch03} & 20.8 \xpm 1.3  \\
\underline{GSI1 + GSI2:}  & (20.7 \xpm 0.9) \\
Seattle(02) \cite{Seatt} & 22.1 \xpm 0.6 \\
Weizmann(03) \cite{Weiz} & 21.2 \xpm 0.7 \\
\   \\
\underline{Average:} & 21.2 \xpm 0.5  \\
{\phantom{$00$}}\\
\hline
\end{tabular}
\end{table}

\begin{table}[hb] 

\caption[TableII]{Extracted slopes of \xs17. An M1 contribution 
to the direct capture data from the 632 resonance has been subtracted, 
see Fig. 4.}\label{tab2}

\begin{tabular}{lllll}

\hline\\[-10pt]
Experiment & Slope (eV-b/MeV) \\
\hline\\[-10pt]
Seattle(02) \cite{Seatt} & 5.1 \xpm \ 0.6 \\
Weizmann(03) \cite{Weiz} & 6.7 \xpm \ 1.2 \\
GSI \cite{Iw99,Sch03} & 9.5 \xpm \ 2.5  \\

{\phantom{$00$}}\\
\hline

\end{tabular}
\end{table}

The Coulomb dissociation of $^8B$ \cite{Bauer} has been suggested as a viable method 
to measure the cross section of the $^7Be(p,\gamma)^8B$ reaction. After the pioneering 
experiment of the RIKEN1 group \cite{Mot94} several experiments 
were carried out at medium energy heavy ion facilities \cite{Kik,Iw99,Dav01,Sch03} 
using a variety of kinematical regions and different experimental techniques. 
While already the data of RIKEN1 suggest a small if not negligible E2 contribution 
\cite{GaiBer} to the Coulomb dissociation of $^8B$, the MSU group claimed to have 
measured a large effect \cite{Dav01} in the measured asymmetry, but no large E2 
contribution was observed in the angular correlation data of GSI \cite{Sch03}. 

In Fig. 2 we show the extracted \S17 from Coulomb dissociation data, 
where the theoretical extrapolation curve used is from Ref. \cite{DB}. 
The large value for the GSI2 result shown in Fig. 2, should not be confused with 
the smaller \S17 stated in the abstract of the paper of the GSI2 group \cite{Sch03}, 
where a different theoretical extrapolation was used. Both large and small values are 
quoted in the paper of the GSI2 group \cite{Sch03}.

We note that while the RIKEN-GSI data were measured with increasingly higher precision, 
reaching the precision of $\pm$5\% or better, the central value of our measurements has 
risen over the years and stabilized just below 21.0 eV-b, see Fig. 2, independently (and long before) 
the direct capture data of the Seattle \cite{Seatt} and Weizmann \cite{Weiz} groups were 
available. The smaller value for \S17 quoted by the MSU group \cite{Dav01} on the other hand, 
is entirely due to their model dependent assumption (and not a measurement) of large E2 
contribution to the Coulomb dissociation of $^8B$. 

Most current global analysis of the \be7pg reaction concentrate on reviewing 
\S17 alone and make the implicit assumption that theoretical 
predictions of the energy dependence of \xs17 agree at low energies, since  
Nuclear Structure effects are assumed to be negligible at low energy. In 
Fig. 3 we show the results of calculations of two similar models \cite{DB,Sch03}. 
As noted in \cite{Sch03} where one of the two theoretical curves is published, the two 
calculations exhibit a significantly different energy dependence, in particular 
the predicted slopes are different. 

An extrapolation of the data of the GSI group using the calculation of Ref. \cite{DB} 
yields \S17 = 20.8 eV-b, a smaller value of 18.6 eV-b is obtained using the potential model 
described in Ref. \cite{Sch03}. The same potential model yields an extrapolated \S17 = 18.1 
eV-b for the Weizmann data \cite{Weiz,Sch03}. Hence one must use caution when extrapolating 
\xs17 to zero energies, or when performing a global analysis of \S17.

We emphasize that thus far available models predict \S17 $\approx$ 1.1 \xs17(200), 
hence add a 10\% extrapolation correction to \xs17 that is now measured with 
3-5\% uncertainty. The confusion on the theoretical side discussed above, makes it 
important to test these theoretical extrapolations at very low energies.
 
The astrophysical cross section factor at low energies (200-1300 keV) exhibit a linear 
dependence on energy, \xs17(E) = \xs17(200) + E x S', where S' is the slope. 
The cross section of \be7pg reaction at low energies is dominated by contributions 
from the s- and d- partial waves \cite{Ham73,Jen}. The predicted slope (S') results from the 
sum of a positive slope (approximately 10 eV-b/MeV) for the d-wave and a slightly negative slope 
(approximately -3 eV-b/MeV) for the s-wave \cite{Jen}. Hence the exact value of the predicted slope 
is model dependent with variation depending on details of the model used. At very low energies 
(below 200 keV) a sharp rise in the s-wave contribution is predicted due to the distortion of 
the plane waves by the Coulomb field. The cross section factor at zero 
energies, \S17, is directly related to the Asymptotic Normalization Coefficient (ANC) 
of the wave function, which is related to the measured spectroscopic factor.

A large(r) d-wave component to the cross section of the \be7pg increases the predicted slope, 
and in this sense the situation is very reminiscent of the extrapolation of the cross 
section of the $^2H(^2H,\gamma)^4He$ \cite{DDG}, where a dominance of one partial 
wave (in this case the d-wave) was assumed by theory. Later it was shown that a very minute 
effect in the Nuclear Structure of $^4He$, changed the predicted extrapolated value 
by a factor of 32 \cite{DDG}. In the case of the $^2H(^2H,\gamma)^4He$ reaction a neglect 
of a small non d-wave contribution to the theoretical extrapolation, drastically altered the 
prediction at energies below  100 keV. In the case of the \be7pg reaction a dominance of one 
partial wave below 100 keV (in this case the s-wave) is also calculated by most theories and this analogy calls 
for extra caution in determining the d-wave or any other non s-wave component when extrapolating 
\S17 with an accuracy of \xpm 5\% or better. A measurement of the slope may in fact yield the exact 
contribution of the d-wave and constraint the theory.

In view of the conflicting data shown in Fig. 1, and in the absence of understanding 
(and correction) of the systematical differences one must make a choice 
for data used to deduce \S17. Similarly one also must make a choice for theoretical 
extrapolation. If we use only the high precision and consistent data of the Seattle, 
Weizmann, GSI1, and GSI2 \cite{Seatt,Weiz,Iw99,Sch03} with the extrapolation procedure 
of Ref. \cite{DB}, we quote: \S17 = 21.2 \xpm 0.5 eV-b, with $\chi^2/\nu$ = 1.06, as shown 
in Table I.

However, as shown in Fig. 4, these high precision data measured over the energy range of 
200 - 1300 keV exhibit different slopes, as listed in Table II. Note that at low energies the 
Weizmann data \cite{Weiz} are systematically below the Seattle data \cite{Seatt}, by up to 2 
sigma. But at higher energies they are in agreement, indicating a systematic difference of the 
measured slope as listed in Table II.

When comparing theoretical calculations to measured data, one multiplies the 
calculations by a normalization factor so as to obtain the measured  
absolute value of the cross section factor, as shown in Fig. 3. In doing so 
one alters the calculated slope (d-wave mixture) of the theory by the same normalization 
factor, thus calling into question the very use of the corrected theory for extrapolating 
to zero energy the very same data that is used to correct the theory. Such a circular 
process can only be viewed as a consistency check and not a determination of facts. 
None-the-less the resulting normalized slope calculated in Ref \cite{DB} 
is 5.0 eV-b/MeV, note the original slope calculated by Descouvemont and Baye is 6.8 eV-b/MeV 
\cite{DB}. A more recent calculations by the same author \cite{Des} exhibit a slightly larger 
slope. More importantly the new calculations reproduce the absolute value of the cross section and 
there is no need to renormalize the theoretical curve to the data.  The non-normalized slope 
of the new calculations is 7.5 eV-b/Mev \cite{Des}. The normalized slope calculated by Typel  
\cite{Sch03} is 8.3 eV-b/MeV and indeed close to the slope of the new calculations \cite{Des}. 
The dispersion of normalized calculated slopes is similar to the dispersion of measured slopes 
listed in Table II. The current situation where one can not rule out either theory or data 
and the confusion with the value of the measured and predicted slopes (S'), is clearly 
unsatisfactory as it lead to a wide spread of extrapolated \S17 values.

For example, as discussed above, an extrapolation of the data included in Table I with the 
Typel theoretical curve yield smaller \S17 values, as low as 18.1 eV-b \cite{Sch03}, see  
Fig. 3. Clearly the measured (as well as the calculated) slopes are sufficiently 
different to preclude extrapolating \S17 with high accuracy (\xpm 5\%), and we must include the 
possible lower extrapolated value. This adds an error of +0.0 -3.0 eV-b due to extrapolation and
we may quote:

\begin{center}
\S17 = 21.2 \xpm 0.5 \ $^{+0.0}_{-3.0}$ (extrap) eV-b
\end{center}

We conclude that a high precision measurement of the slope of \xs17 
extending to very low energies (below 100 keV) is needed in order to 
quote \S17 with an uncertainty of \xpm 5\% or better. This is a formidable 
task since it requires high precision measurement of all associated multiplicative 
factors that are relevant for the measurement. 

A high precision measurement of the slope is possible when measuring the ratio of two 
cross sections which involve the same experimental multiplicative factors; e.g. the ratio of 
the cross section of the \be7pg reaction relative to the elastic (Coulomb) scattering cross 
section. Such an experiment can be performed (only) with the use of \xbe7 beams. Indeed a proposal 
for such an experiment at CERN-ISOLDE \cite{ISOLDE} was considered. 

Clearly in the absence of such data on the slope (S') and the discussion above concerning the 
extrapolation to zero energy, the situation with \S17 will remain confused and \S17 can not be 
quoted with the accuracy of \xpm 5\% or better, and a down systematic uncertainty of -15\% must 
be included.

The author thanks John Bahcall and Pierre Descouvemont for sharing their theoretical 
results, Refs \cite{Bahcall,Des} respectively, prior to publication.


\begin{thebibliography}{9}

\bibitem{SNO} S.N. Ahmad {\em et al.}; nucl-ex/0309004.

\bibitem{RoadMap} J.N. Bahcall and C. Pena-Garay; JHEP {\bf 11}(2003)004.

\bibitem{SSM} J.N. Bahcall, M.H. Pinsonneault, and S. Basu; ApJ {\bf 555}(2001)990.

\bibitem{Bah03} J.N. Bahcall, N.C. Gonzalez-Garcia, C. Pena-Garay; JHEP 
 {\bf 02}(2003)009.

\bibitem{Ham01} F. Hammache {\em et al.}; Phys. Rev. Lett. {\bf 86}(2001)3985.

\bibitem{Str01} F. Strieder {\em et al.}; Nucl. Phys. {\bf A696}(2001)219.

\bibitem{Seatt} A.R. Junghans {\em et al.}; Phys. Rev. Lett {\bf 88}(2002)041101,
 ibid Phys. Rev. {\bf C68}(2003)065803, ibid nucl-ex/0311012.

\bibitem{Weiz} L.T. Baby {\it et al.}, Phys. Rev. Lett. {\bf 90}(2003)022501, ibid
 Phys. Rev. {\bf C67}(2003)065805, ER {\bf C69}(2004)019902(E).

\bibitem{Iw99} N. Iwasa {\em et al.}; Phys. Rev. Lett. {\bf 83}(1999)2910.

\bibitem{Sch03} F. Schumann {\em et al.}; Phys. Rev. Lett. {\bf 90}(2003)232501.

\bibitem{Dav01} B. Davids {\em et al.}; Phys. Rev. Lett. {\bf 86}(2001)2750.

\bibitem{Adel} E.G. Adelberger {\em et al.}; Rev. of Modern Phys. {\bf 70}(1998)1265.

\bibitem{Ham} R.G.H. Robertson {\em et al.}; Phys. Rev. {\bf C27}(1983)27.

\bibitem{Bauer} G. Baur, C.A. Bertulani, and H. Rebel;
 Nucl. Phys. {\bf A458}(1986)188.

\bibitem{Mot94} T. Motobayashi {\em et al.}; Phys. Rev. Lett. {\bf 73}(1994)2680.

\bibitem{Kik} T. Kikuchi {\em et al.}; Phys. Lett. {\bf B391}(1997)261, 
 ibid E. Phys. J. {\bf A3}(1998)213.

\bibitem{GaiBer} M. Gai and C.A. Bertulani; Phys. Rev. {\bf C52}(1995)1706.

\bibitem{DB} P. Descouvemont and D. Baye; Nucl. Phys. {\bf A567}(1994)341.

\bibitem{Bahcall} J.N. Bahcall and M. H. Pinsonneault; astro-ph/0402114.

\bibitem{Ham73} R.G.H. Robertson; Phys. Rev. {\bf C7}(1973)543.

\bibitem{Jen} B.K. Jennings, S. Karataglidis, and T.D. Shoppa; Phys. Rev. 
 {\bf C58}(1998)3711. 

\bibitem{DDG} C.A. Barnes, K.H. Chung, T.R. Donoghue, C. Rolfs,
and J. Kammeraad; Phys. Lett. {\bf B197}(1987)315.

\bibitem{Des} P. Descouvemont, and M. Dufour; Contribution, 8th Int. Conf. on Clustering, 
November 24-29, 2003, Nara, Japan, page 77.

\bibitem{ISOLDE} M. Gai, M. Hass and Th. Delbar, for the UConn-Weizmann-LLN-ISOLDE  
collaboration;  Letter of Intent to INTC, I37 INTC 2001-007.

\end{thebibliography}
\end{document}